\documentclass{my-ws-p10x7}

\begin{document}

\title{
\hspace*{-0.5cm}
{\rm \mbox{(PITHA 00/13, hep-ph/0007256, July 2000)}}\\[0.7cm]
\boldmath 
QCD factorization for $B\to\pi K$ decays
\unboldmath}

\author{
\vspace*{-0.2cm} M.~Beneke}

\address{
Institut f\"ur Theoretische Physik E, RWTH Aachen, 
D - 52056 Aachen, Germany\\E-mail: mbeneke@physik.rwth-aachen.de}

\author{\vspace*{-0.2cm} G.~Buchalla}

\address{Theory Division, CERN, CH-1211 Geneva 23, Switzerland}  

\author{\vspace*{-0.2cm} M.~Neubert}

\address{Newman Laboratory of Nuclear Studies, 
Cornell University, Ithaca, NY 14853,  USA}  

\author{\vspace*{-0.2cm} C.T.~Sachrajda}

\address{Dept. of Physics and Astronomy, University
  of Southampton, Southampton SO17 1BJ, UK}  

\twocolumn[\maketitle\abstract{
We examine some consequences of the QCD factorization approach to 
non-leptonic $B$ decays into $\pi K$ and $\pi \pi$ 
final states, including a set of enhanced power corrections. Among 
the robust predictions of the approach we find small 
strong-interaction phases (with one notable exception) and a 
pattern of CP-averaged branching 
fractions, which in some cases differ significantly from the current  
central values reported by the CLEO Collaboration.}]

\section{Introduction}

\vspace*{-0.2cm}
The observation of $B$ decays into $\pi K$ and $\pi \pi$ final 
states \cite{} has resulted in a large amount of theoretical and 
phenomenological work that attempts to interpret these 
observations in terms of the factorization approximation (FA), or in 
terms of general parameterizations of the decay amplitudes. 
A detailed understanding of these amplitudes would help us to pin down 
the value of the CKM angle $\gamma$ using only data on CP-averaged
branching fractions. Theoretical work on the heavy-quark limit has 
justified the FA as a useful starting 
point\cite{BBNS99,BBNS00}, but predicts important and computable 
corrections. Here we discuss the most important consequences of 
this approach for the $\pi K$ and $\pi \pi$ final 
states.\footnote{Contribution to ICHEP2000, July 27 - August 2, Osaka, 
Japan (talk presented by M.~Beneke).}

To leading order in an expansion in powers of $\Lambda_{\rm QCD}/m_b$, 
the $B\to \pi K$ matrix elements obey the factorization 
formula
\begin{eqnarray}\label{fact}
   \langle\pi K|Q_i|B\rangle
   &=& f_+^{B\to\pi}(0)\,f_K\,T_{K,i}^{\rm I}*\Phi_K \nonumber\\
   &&\hspace*{-1.5cm}\mbox{}
    + f_+^{B\to K}(0)\,f_\pi\,T_{\pi,i}^{\rm I}*\Phi_\pi \\ 
   &&\hspace*{-1.5cm}\mbox{}
    + f_B f_K f_\pi\,T_i^{\rm II}*\Phi_B*\Phi_K*\Phi_\pi \,,
    \nonumber
\end{eqnarray}
where $Q_i$ is an operator in the weak effective Hamiltonian, 
$f_+^{B\to M}(0)$ are semi-leptonic form factors of a vector
current evaluated at $q^2=0$, $\Phi_M$ are leading-twist
light-cone distribution amplitudes, and the $*$-products imply an
integration over the light-cone momentum fractions of the 
constituent quarks inside the mesons. When the hard-scattering 
functions $T$ are evaluated to order $\alpha_s^0$, Eq.~(\ref{fact}) 
reduces to the conventional FA. The 
subsequent results are based on kernels including all corrections 
of order $\alpha_s$. A detailed justification of (\ref{fact}) is 
given in Ref.~\cite{BBNS00}. Compared to our previous discussion 
of $\pi\pi$ final states\cite{BBNS99} the present analysis 
incorporates three new ingredients: 

i) the matrix elements of electroweak (EW) penguin operators 
(for $\pi K$ modes); 

ii) hard-scattering kernels for general, asymmetric light-cone 
distributions; 

iii) the complete set of ``chirally enhanced'' $1/m_b$ 
corrections.\cite{BBNS99}

\noindent
The second and third items have not been considered in 
other\cite{DZ00} generalizations of Ref.~\cite{BBNS99} to the $\pi K$ 
final states. The third one, in particular, is essential for estimating 
some of the theoretical uncertainties of the approach. 


We now briefly present the input to our calculations. 
Following Ref.~\cite{BBNS99},
we obtained the coefficients $a_i(\pi K)$ of 
the effective factorized transition operator defined analogously to the 
case of $\pi\pi$ final states, but augmented by
coefficients $a_{7-10}(\pi K)$ related to EW penguin 
operators and electro-magnetic penguin contractions of current--current 
and QCD penguin operators. A sensible 
implementation of QCD corrections to EW penguin matrix elements
implies that one departs from the usual renormalization-group 
counting, in which the initial condition for EW penguin coefficients is 
treated as a next-to-leading order (NLO) 
effect. Our NLO initial condition hence includes 
the $\alpha_s$ corrections computed in Ref.~\cite{BGH00}. 

Chirally enhanced corrections arise from twist-3 two-particle 
light-cone distribution amplitudes, whose normalization involves the 
quark condensate. The relevant parameter, 
$2\mu_\pi/m_b = -4\langle \bar{q}q\rangle/(f_\pi^2 m_b)$, is formally 
of order $\Lambda_{\rm QCD}/m_b$, but large numerically. The
coefficients $a_6$ and $a_8$ are multiplied by this parameter.
There are also additional chirally enhanced corrections to the 
spectator-interaction term in (\ref{fact}), which turn out to be the 
more important effect. In both cases, these corrections involve 
logarithmically divergent integrals, which violate factorization. For
instance, for matrix elements of $V-A$ operators the hard spectator 
interaction is now proportional to ($\bar u\equiv 1-u$)
\[
   \int_0^1 \!\frac{du}{\bar{u}} \frac{dv}{\bar{v}}\, 
   \Phi_K(u)\left(\Phi_\pi(v)+\frac{2\mu_\pi}{m_b} \frac{\bar{u}}{u}
   \right) 
\]
when the spectator quark goes to the pion. (Here we used that the 
twist-3 distribution amplitudes can be taken to be the asymptotic
ones when one neglects twist-3 corrections without the chiral 
enhancement.) The divergence of the $v$-integral in the second term
as $\bar v\to 0$ implies that it is dominated by soft gluon
exchange between the spectator quark and the quarks that form the 
kaon. We therefore treat the divergent integral $X=\int_0^1(dv/\bar v)$
as an unknown parameter (different for the penguin and hard scattering
contributions), which may in principle be complex owing 
to soft rescattering in higher orders. In our numerical analysis we
set $X=\ln(m_B/0.35\,\mbox{GeV})+r$, where $r$ is chosen randomly 
inside a circle in the complex plane of radius 3 (``realistic'') or
6 (``conservative''). 
Our results also depend on the $B$-meson parameter\cite{BBNS99} $\lambda_B$,
which we vary between 0.2 and 0.5\,GeV. Finally, there is in some 
cases a non-negligible dependence of the coefficients $a_i(\pi K)$
on the renormalization scale, which we vary between $m_b/2$ and 
$2m_b$.

\vspace*{-0.2cm}
\section{Results}

\vspace*{-0.2cm}
We take $|V_{ub}/V_{cb}|=0.085$ and 
$m_s(2\,\mbox{GeV})=110\,$MeV as fixed input, noting that ultimately
the ratio $|V_{ub}/V_{cb}|$, along with the CP-violating phase 
$\gamma=\mbox{arg}(V_{ub}^*)$, might be extracted from a simultaneous fit
to the $B\to\pi K$ and $B\to\pi\pi$ decay rates.

\vspace*{-0.3cm}
\subsection{$SU(3)$ breaking}

\vspace*{-0.1cm}
Bounds\cite{FM98,NR98} on the CKM angle $\gamma$ derived from ratios of 
$\pi K$ branching fractions, as well as the determination of $\gamma$ 
using the method of Ref.~\cite{NR}, rely on an estimate of $SU(3)$ 
flavour-symmetry violations. We find that ``non-factorizable'' 
$SU(3)$-breaking effects (i.e., effects not accounted for by the 
different decay constants and form factors of pions and kaons in the 
conventional FA) do not exceed a few percent at leading power.

\vspace*{-0.3cm}
\subsection{Amplitude parameters}

\vspace*{-0.1cm}
The approach discussed here allows us to obtain the decay amplitudes 
for the $\pi\pi$ and $\pi K$ final states in terms of the form factors 
and the light-cone distribution amplitudes. The $\pi^0\pi^0$ 
final state is very poorly predicted and will not be discussed here. 
We write 
\begin{displaymath}
{\cal A}(B^0\to\pi^+\pi^-) = T\,[e^{i\gamma}+(P/T)_{\pi\pi}] 
\end{displaymath}
and parameterize the $\pi K$ amplitudes by\cite{NR98}
\begin{eqnarray}\label{para}
   {\cal A}(B^+\to\pi^+ K^0) &=& P \left( 1 - \varepsilon_a\,
    e^{i\eta} e^{i\gamma} \right), \nonumber\\
   - \sqrt2\,{\cal A}(B^+\to\pi^0 K^+) &=& P \Big[ 1
    - \varepsilon_a\,e^{i\eta} e^{i\gamma} \nonumber\\ 
   &&\hspace*{-2cm}\mbox{}
    - \varepsilon_{3/2}\,e^{i\phi} (e^{i\gamma} - q\,e^{i\omega})
    \Big],\\
   - {\cal A}(B^0\to\pi^- K^+) &=& P \Big[ 1
    - \varepsilon_a\,e^{i\eta} e^{i\gamma} \nonumber\\ 
   &&\hspace*{-2cm}\mbox{}
   - \varepsilon_T\,e^{i\phi_T} (e^{i\gamma} - q_C\,e^{i\omega_C})
    \Big] , \nonumber 
\end{eqnarray}
and $\sqrt2\,{\cal A}(B^0\to\pi^0 K^0) = {\cal A}(B^+\to\pi^+ K^0)
+ \sqrt2\,{\cal A}(B^+\to\pi^0 K^+) - {\cal A}(B^0\to\pi^- K^+)$. 
Table~\ref{tab1} summarizes the numerical values for the  
amplitude parameters for the conservative variation of $X$,
and variation of the other parameters as explained above.
The LO results correspond to the conventional FA at the fixed
scale $\mu=m_b$. They are strongly scale dependent. In comparison, the 
scale-dependence of the NLO result is small, with the exception 
of $q_C\,e^{i \omega_C}$. 
One must keep in mind that the ranges may overestimate the 
true uncertainty, since the parameter $X$ may ultimately 
be constrained from a subset of branching fractions. This is true 
in particular for the quantity $\varepsilon_{3/2}$ in
Table~\ref{tab1}, which can be extracted from data.\cite{NR98}

\begin{table}
\caption{Parameters for the $B\to\pi K$ amplitudes as defined in 
(\protect\ref{para}), for conservative variation of all input 
parameters (see text).} 
\label{tab1}

\begin{center}
\begin{tabular}{|c|c|c|} 
\hline 
 \raisebox{0pt}[12pt][6pt]{} & 
\raisebox{0pt}[12pt][6pt]{Range, NLO} & 
 \raisebox{0pt}[12pt][6pt]{LO}  \\
 \hline
 \raisebox{0pt}[12pt][6pt]{$-\varepsilon_a\,e^{i\eta}$} & 
 \raisebox{0pt}[12pt][6pt]{$(0.017\mbox{--}0.020)\,e^{i\,[13,21]^\circ}$}
 & \raisebox{0pt}[12pt][6pt]{$0.02$}\\
\hline
 \raisebox{0pt}[12pt][6pt]{$\varepsilon_{3/2}\,e^{i\phi}$} & 
 \raisebox{0pt}[12pt][6pt]{$(0.20\mbox{--}0.38)\,e^{i\,[-30,7]^\circ}$}
 & \raisebox{0pt}[12pt][6pt]{$0.36$}\\
\hline
 \raisebox{0pt}[12pt][6pt]{$q\,e^{i\omega}$} & 
 \raisebox{0pt}[12pt][6pt]{$(0.53\mbox{--}0.63)\,e^{i\,[-7,3]^\circ}$}
 & \raisebox{0pt}[12pt][6pt]{$0.64$}\\
\hline
 \raisebox{0pt}[12pt][6pt]{$\varepsilon_T\,e^{i\phi_T}$} & 
 \raisebox{0pt}[12pt][6pt]{$(0.20\mbox{--}0.29)\,e^{i\,[-19,3]^\circ}$}
 & \raisebox{0pt}[12pt][6pt]{$0.33$}\\
\hline
 \raisebox{0pt}[12pt][6pt]{$q_C\,e^{i\omega_C}$} & 
 \raisebox{0pt}[12pt][6pt]{$(0.00\mbox{--}0.22)\,e^{i\,[-180,180]^\circ}$}
 & \raisebox{0pt}[12pt][6pt]{$0.06$}\\
\hline
 \raisebox{0pt}[12pt][6pt]{$(P/T)_{\pi\pi}$} & 
 \raisebox{0pt}[12pt][6pt]{$(0.19\mbox{--}0.29)\,e^{i\,[-1,23]^\circ}$}
 & \raisebox{0pt}[12pt][6pt]{$0.16$}\\
\hline
\end{tabular}
\end{center}
\vspace*{-0.3cm}
\end{table}

\vspace*{-0.3cm}
\subsection{Ratios of CP-averaged rates}

\begin{figure}
\hspace*{-0.7cm}
\epsfxsize180pt
\figurebox{120pt}{160pt}{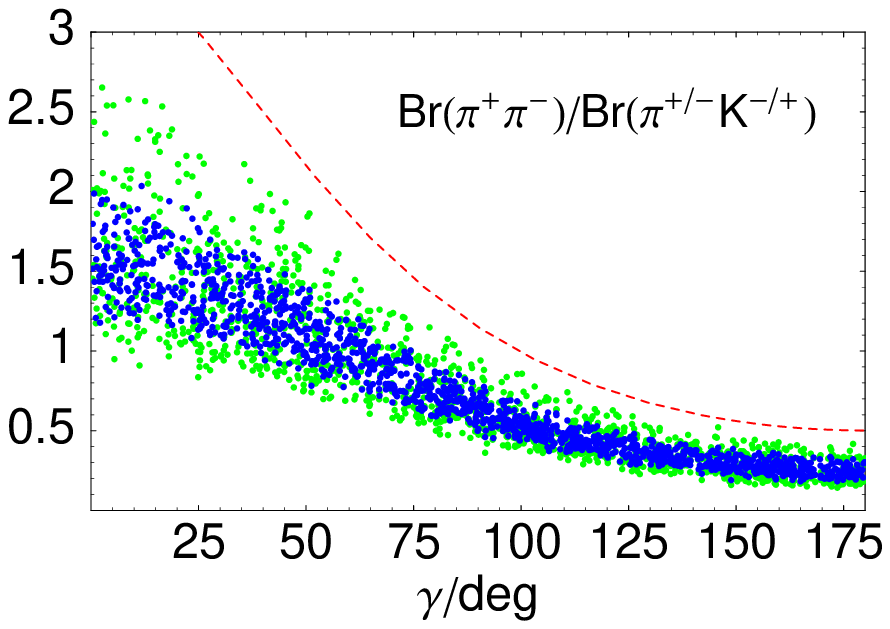}
\vspace*{-0.1cm}
\epsfxsize180pt
\figurebox{120pt}{160pt}{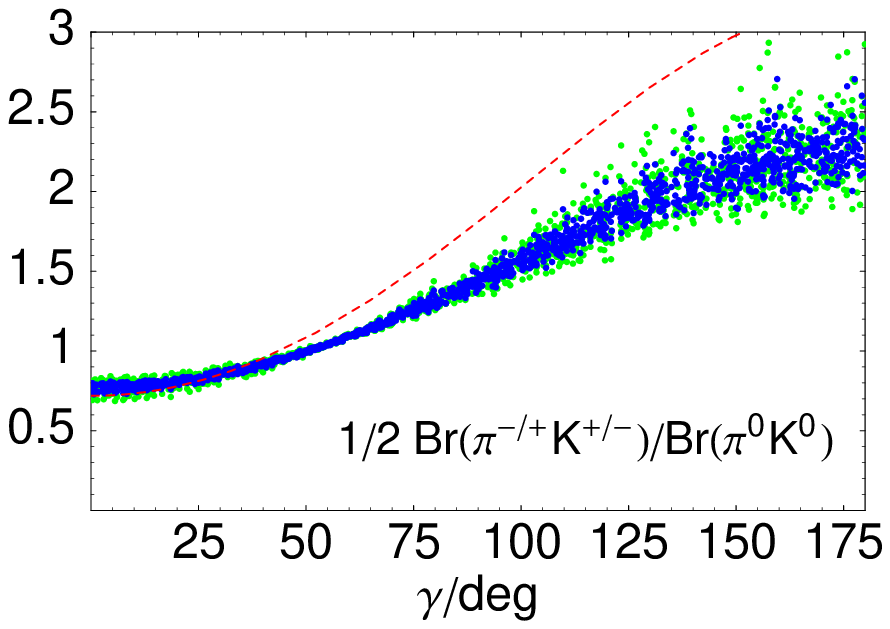}
\vspace*{-0.1cm}
\epsfxsize180pt
\figurebox{120pt}{160pt}{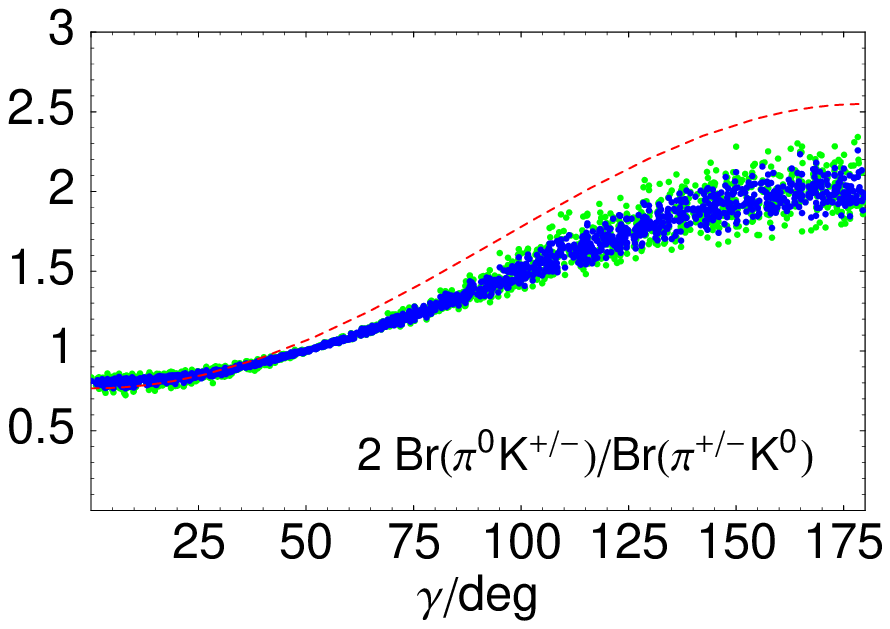}
\vspace*{-0.1cm}
\epsfxsize180pt
\figurebox{120pt}{160pt}{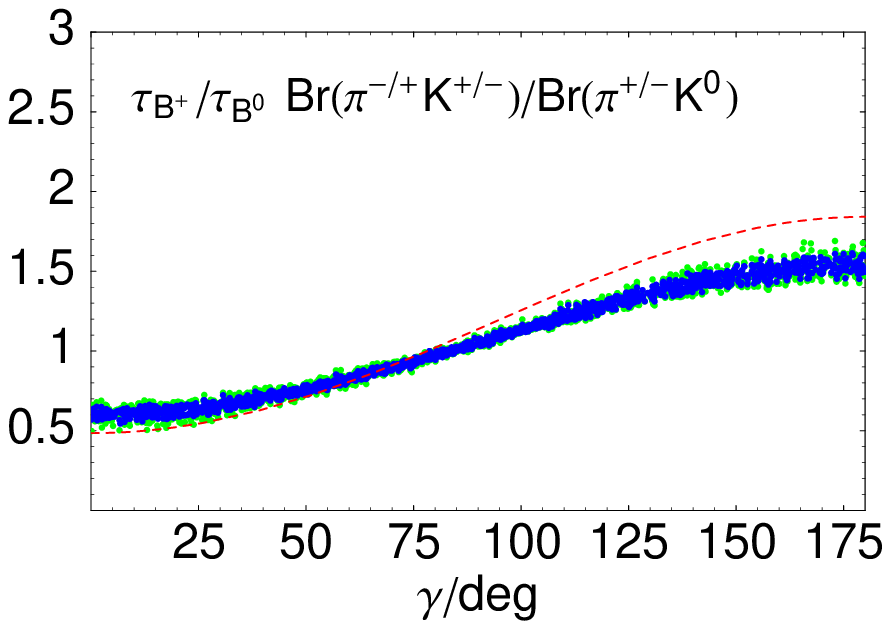}
\vspace*{-0.1cm}
\epsfxsize180pt
\figurebox{120pt}{160pt}{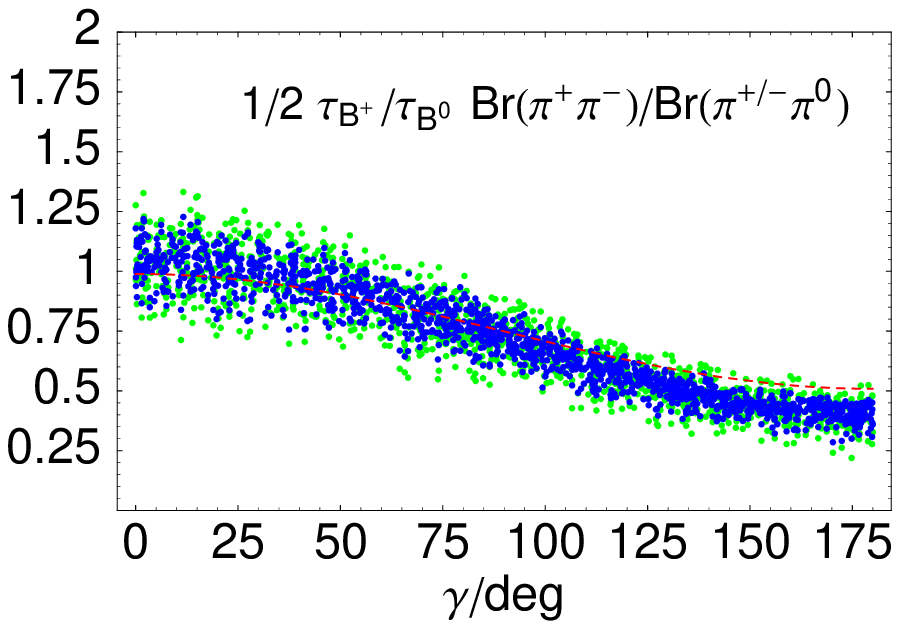}
\vspace*{-0.2cm}
\caption{Ratios of CP-averaged $B\to\pi K$ and $\pi\pi$ decay rates. 
The scattered points cover a realistic (dark) and conservative (light)
variation of input parameters. The dashed curve is the LO result, 
corresponding to conventional factorization.}
\label{fig1}
\end{figure}

\vspace*{-0.1cm}
Since the form factor $f_+(0)$ is not well known, we consider here
only ratios of CP-averaged branching ratios, discarding the 
$\pi^0\pi^0$ final state. We 
display these as functions of the CKM angle $\gamma$ in 
Fig.~\ref{fig1}. 


Table~\ref{tab1} shows that the corrections with respect to the 
conventional FA are significant (and 
important to reduce the re\-nor\-malization-scale dependence). Despite 
this fact, the {\em qualitative } pattern that emerges for the set of 
$\pi K$ and $\pi \pi$ decay modes is similar to that in conventional 
factorization. In particular, the penguin--tree interference is 
constructive (destructive) in $B\to\pi^+\pi^-$ ($B\to\pi^- K^+$)
decays if $\gamma<90^\circ$. Taking the currently favoured range 
$\gamma=(60\pm 20)^\circ$, we find the following robust predictions:
\begin{eqnarray}
   \frac{\mbox{Br}(\pi^+\pi^-)}{\mbox{Br}(\pi^\mp K^\pm)}
   &=& 0.5\mbox{--}1.9 \quad [0.25\pm 0.10] \nonumber\\
   \frac{\mbox{Br}(\pi^\mp K^\pm)}{2\mbox{Br}(\pi^0 K^0)}
   &=& 0.9\mbox{--}1.4 \quad [0.59\pm 0.27] \nonumber\\
   \frac{2\mbox{Br}(\pi^0 K^\pm)}{\mbox{Br}(\pi^\pm K^0)}
   &=& 0.9\mbox{--}1.3 \quad [1.27\pm 0.47] \nonumber\\
   \frac{\tau_{B^+}}{\tau_{B^0}}\,
   \frac{\mbox{Br}(\pi^\mp K^\pm)}{\mbox{Br}(\pi^\pm K^0)}
   &=& 0.6\mbox{--}1.0 \quad [1.00\pm 0.30] \nonumber
\end{eqnarray}
The first ratio is in striking disagreement with 
current CLEO data\cite{C00} (square brackets). 
The near equality of the second and third ratios is a 
result of isospin symmetry.\cite{NR98} We find
$\mbox{Br}(B\to\pi^0 K^0)=(4.5\pm 2.5)\times 10^{-6}\, 
(V_{cb}/0.039)^2 (f^{B\to\pi}_+(0)/0.3)^2$
almost 
independently of $\gamma$. This is three time smaller than the central
value reported by CLEO.

\vspace*{-0.3cm}
\subsection{CP asymmetry in $B\to \pi^+\pi^-$ decay}

\begin{figure}
\epsfxsize190pt
\figurebox{120pt}{160pt}{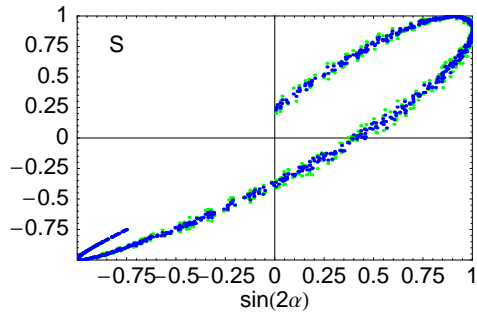}
\vspace*{-0.1cm}
\caption{Mixing-induced CP asymmetry in $B\to \pi^+\pi^-$ decays. 
The lower band refers to values $45^\circ<\alpha<135^\circ$, the
upper one to $\alpha<45^\circ$ (right) or $\alpha>135^\circ$ (left). 
We assume $\alpha,\beta,\gamma\in [0,\pi]$.}
\label{fig2}
\vspace*{-0.5cm}
\end{figure}

\vspace*{-0.1cm}
The stability of the prediction for the $\pi^+\pi^-$ amplitude 
suggests that the CKM angle $\alpha$ can be extracted from 
the time-dependent mixing-induced CP asymmetry in this decay mode, 
without using isospin analysis. Fig.~\ref{fig2} displays the 
coefficient $S$ of $-\sin(\Delta M_{B_d} t)$ as a function of 
$\sin(2\alpha)$ for $\sin(2\beta)=0.75$, which may be compared with 
the result in Ref.~\cite{B99}. For some values of $S$ there
is a two-fold ambiguity (assuming all angles are between 
$0^\circ$ and $180^\circ$). A consistency check of the approach 
could be obtained, in principle, from the coefficient of the 
$\cos(\Delta m_{B_d} t)$ term.


\vspace*{-0.2cm}
\section{Conclusions}

\vspace*{-0.2cm}
We have examined some of the consequences of the QCD factorization 
approach to $B$ decays into $\pi K$ and $\pi \pi$ final 
states, leaving a detailed discussion to a subsequent publication. 
Here we have focused on robust predictions for ratios of CP-averaged
decay rates.
Our result for the ratio of the $B\to\pi^+\pi^-$
and $B\to\pi^\mp K^\pm$ decay rates is in disagreement with the 
current experimental value, unless the weak phase $\gamma$ were 
significantly larger than $90^\circ$. 

\end{document}